\documentclass{cjaa}                    
\usepackage{graphicx}                   
\def \hcm {\hbox {\ifmmode $ cm$^{-2}\else cm$^{-2}$\fi}}

\def\approxgt{\mathrel{\hbox{\rlap{\lower.55ex \hbox {$\sim$}}
        \kern-.3em \raise.4ex \hbox{$>$}}}}
\def\approxlt{\mathrel{\hbox{\rlap{\lower.55ex \hbox {$\sim$}}
        \kern-.3em \raise.4ex \hbox{$<$}}}}

\begin{document}
   \title{Measurement of Mass and Spin of Black Holes with QPOs
}
   \author{B. Aschenbach
      \inst{}\mailto{}
                }
   \offprints{B. Aschenbach}                   
   \institute{Max-Planck-Institut f\"ur Extraterrestrische Physik,
      P.O. Box 1312, Garching, 85741, Germany\\
             \email{bra@mpe.mpg.de}
          }
   \date{Received~~2007 month day; accepted~~2007~~month day}
   \abstract{There are now four low mass X-ray binaries with black holes which show twin resonant-like HFQPOs. Similar QPOs 
might have been found in Sgr A$\sp *$. I review the power spectral density distributions of the three X-ray flares and the six 
NIR flares published 
 for  Sgr A$\sp *$ so far, in order to look for more similarities than just the frequencies 
between the microquasar black holes and Sgr A$\sp *$.
The three X-ray flares of Sgr A$\sp *$ are re-analysed in an identical way and white noise probabilities from their power density 
distributions are given for the periods reported around $\sim$ 1100 s. 
Progress of the resonant theory using the anomalous orbital velocity effect is summarized.   
\keywords{Galaxy: center - X-rays: general - black hole physics - X-rays: individuals - Sgr A$\sp*$}
   }
   \authorrunning{B. Aschenbach}            
   \titlerunning{Black hole QPOs}  
\maketitle
\section{Introduction}           
\label{sect:intro}
Quasi-periodic oscillations (QPOs) are believed to arise from variations of the accretion flow around 
compact objects, i.e. white dwarfs, neutron stars and black holes. As far as black holes are concerned, 
Remillard and McClintock (2006) have compiled a list of a total of 40 sources  
which show up in galactic X-ray binaries, 20 of which show a dark companion with a mass largely exceeding the 
mass of a neutron star but apparently limited to about 18 solar masses. Out of the 20 sources with established 
dynamically measured masses 16 sources show low frequency QPOs, whereas high frequency QPOs (HFQPOs, $\nu$ $>$ 10 Hz) have been 
detected in 8 sources. 

\section{The recent past of the resonant QPOs}

A remarkable event was the detection of twin HFQPOs which started with GRO J1655-40 and 
GRS 1915+105  by Strohmayer (2001a, 2001b). Before this discovery Klu\'zniak \& Abramowicz (2001) pointed out 
that such twin QPOs could be produced as a non-linear 1:2 or 1:3 resonance between orbital and radial 
epicyclic motion. 
The formulas for the epicyclic frequencies can be found, for instance, in the papers by Aliev and Galtsov (1981) or 
Nowak and Lehr (1998).
In a subsequent paper Abramowicz \& Klu\'zniak (2001) noted that the two frequencies (450 Hz and 300 
Hz) then known for GRO J1655-40 come in a 3:2 ratio, in support of their model. Given the dynamically measured 
mass, they estimated that the black hole was spinning with a Kerr parameter of 0.2$< \it a <$ 0.67, using a resonance 
between orbital and radial epicyclic motion.

About a year later Abramowicz et al. (2002) started to consider the 3:2 twin frequencies to arise from a parametric 
resonance between the vertical and radial epicyclic motions, which from then on was studied in a series  of  papers, 
 e.g., Abramowicz et al. (2003, 2004). 
McClintock \& Remillard (2004) noted that the upper frequency of the three 3:2 HFQPO black holes with well 
established dynamical masses, i.e. GRO J1655-40, XTE J1550-564 and GRS 1915+105, was consistent with the frrequencies  
scaling inversely proportional to the black hole mass. Later on this was used as supporting evidence that HFQPOs 
were solely due to general relativistic effects. 

If these relativistic effects are associated with resonances 
between epicyclic motions, this statement is, strictly speaking, incorrect. For instance, the association 
of the 3:2 twin frequencies with a parametric resonance between the vertical and the radial 
epicyclic frequencies provides a relationship between the orbital resonance radius r$\sb{32}$ and the spin {\it a} 
of the black hole. For determining the black hole mass any one of the  $\{$r$\sb{32}, {\it a}$$\}$ pairs  
can be used leaving an uncertainty of about a factor of 3.4 in the determination of the black hole mass. 
Over a short range of 
black hole masses involved like stellar size mass black holes a well constrained 1/M relation would imply 
that the different black hole objects have the same resonance radius and the same spin.  
Furthermore, if epicyclic 3:2 parametric resonances produce the observed frequencies  
 we are dealing with motion in the innermost region of the accretion disk of less than 10.8 gravitational radii, or 
5.4 Schwarzschild radii. 
 
If the HFQOs are associated with the epicyclic frequencies and are launched at the same orbital radius, mass and spin can be 
uniquely determined by measuring 
three independent frequencies, because there are just three unknowns, which are the orbital radius, the spin and the mass of the 
black hole. General relativity provides three different frequencies, which are the orbital (Kepler) frequency and the 
vertical and the radial epicyclic frequencies. For the 3:2 twins the latter two frequencies are occupied. So one has to look 
for a Kepler frequency. General relativity shows that the Kepler frequency exceeds the vertical (the upper) epicyclic frequency 
by a factor of up to about three close to the innermost stable circular orbit, changing with the spin.  
Appearingly, none of the three microquasars shows any evidence for a frequency higher than the vertical epicyclic frequency. 
This could be interpreted in a way that Keplerian motion of clumps or blobs is negligible as far HFQPO power 
is concerned.

There is the possibility that the Kepler frequency manifests itself in the Lense-Thirring precession frequency, 
which is the difference between Kepler frequency and vertical epicyclic frequency (Merloni et al. 1999).    
For the twin QPO microquasars, adopting the 3:2 twins to be associated with the vertical and radial 
epicyclic frequencies, the Lense-Thirring frequency is calculated to be at least a factor of 3 lower than the
radial epicyclic frequency, thereby falling in the regime of low frequency QPOs. This factor increases 
dramatically with decreasing spin. Looking at the measured QPOs of the microquasars there appears to be  
no indication of such an additional frequency. This might support the conclusion that the manifestation of a Kepler frequency 
in QPOs is lacking, be it a true Keplerian or the Lense-Thirring frequency.

\section{Sgr A$\sp *$}

In 2003 the discovery of a QPO with a period of 16.8 min in the near-infrared (NIR) from a supermassive black hole, 
i.e. Sgr A$\sp *$ in the center of our 
Galaxy, was published. If interpreted as the Keplerian frequency of the innermost stable circular orbit and adopting 
a mass of 3.6$\times$10$\sp 6$ solar masses indicated by dynamical measurements
the spin of the black hole should exceed $\it a >$ 0.5 (Genzel et al. 2003).         
Shortly after that publication Klu\'zniak et al. (2004) suggested that "it seems quite safe to assume that the HFQPOs are 
closely related to orbital frequencies in Einstein's gravity". 
In their 2004 paper Abramowicz et al. remarked:
"If the 17 min flare period does indeed correspond to the upper
(or lower) of the twin peak QPOs in microquasars, it would be interesting to see whether a 26 minute (or 12 minute)
quasi-periodicity may also be present in the source.
After this paper was completed, Aschenbach et al. (astro-ph/0401589) reported
X-ray QPOs from Sgr A$\sp *$. The claimed periods include 1150 s (19 minutes) and 700 s (12 minutes)."

In that paper, Abramowicz et al. (2004) refer to, Aschenbach et al. (2004) suggest that there are five  
quasi-periods in the X-ray flares and the NIR flares of Sgr A$\sp *$ which are conspicuous because of their power spectral density (psd) 
clearly exceeding the mean white noise level. Apart from their psd value the frequencies were picked because excessive 
psd was observed in each of at least two independent observations within the digital frequency resolution of the Fourier 
spectra. The longest three periods are dated between 701~-~692 s, 1173~-~1117 s and ~2307s-~2307s. 
It was actually the coincidence of similar periods in more than one measurement, which makes these periods interesting 
and worth to be looked for in future measurements. A period of 26 minutes argued for by Abramowicz et al. (2004) was not found.

A major drawback of the Aschenbach et al. (2004) model for the association of frequency and type of 
oscillation with just epicyclic, Kepler and 
Lense-Thirring motions involved, was the necessity to invoke different orbital radii. 
Furthermore, the mass of 2.7$\times$10$\sp 6$ solar masses predicted for Sgr A$\sp *$ by this model 
underestimated the best-sold dynamical mass by $\sim$30$\%$, although it was consistent with the dynamical mass 
published by Schoedel et al. (2002) just two years before. 

In a slightly later paper Aschenbach (2004) reported that the three lowest frequencies come in a 3:2:1 ratio, pushing the idea 
that these were true resonant QPOs, essentially 3:1 and 3:2 resonances between the vertical and radial epicyclic frequencies 
of a black hole of fixed spin created at two different orbital radii with commensurable Kepler frequencies, in analogy to 
the Titius-Bode rule. The only solution for such a commensurability is found for the Kepler frequencies at the two different radii 
to be in a 3:1 ratio. With these commensurability conditions there is only one solution for the spin, i.e. $\it a$ = 0.99616  
and the black hole mass, which scales as 4603.3/$\nu\sb{up}$ for the 3:1 resonance and 3046.2/$\nu\sb{up}$ for 
the 3:2 resonance. The mass is given in units of the solar mass. 
$\nu\sb{up}$ is the uppermost frequency in either case and the solution for the resonance orbits is  
 1.546 and 3.919 gravitational radii, respectively. Applied to Sgr A$\sp *$ and the twin QPO microquasars the masses predicted by this 
model are in acceptable agreement with the dynamically measured masses and their uncertainties.  
This has been the first model ever, which predicts mass and spin of a black hole from frequency measurements without any further 
assumption.   

Undoubtedly, the periods suggested have to be confirmed, and undoubtedly,  
the model has to be substantiated by the relevant physics. 
But when searching for the relevant periods, a few things are suggested to be kept in mind, which have been 
taught by 
microquasars studies. 
 
 QPOs are not strictly periodic; the frequency observed in  a single QPO might change from observation to observation 
in a low signal/noise environment. 

QPOs in general have a coherence or quality factor Q~=~$\nu$/{$\Delta\nu\sb{FWHM}$} 
which is basically the inverse 
of the relative bandwidth. For instance, the 450 Hz HFQPO in GRO J1655-40 has a fairly high coherence with Q$\sim$11 or a bandwidth 
of $\pm$4.5\%\ (Strohmayer 2001b). Usually Q is less than this for HFQPOs (Remillard \&\ McClintock 2006).

Twin 
HFQPOs do not show up necessarily  
simultaneous, but in separate observations. HFQPOs seem to be coupled to the X-ray state: "All of the strong detections ... occur 
in the steep power law state (SPL)" (Remillard \&\ McClintock 2006). I note that the first three bright X-ray flares of 
October 2000, October 2002 and August 2004 of Sgr A$\sp *$ show up in different brightness and spectral states, which follow the 
microquasar appearance. The first and the third flare are much less bright than the second flare, which, on the other hand, 
 has a much steeper power law spectrum. This is reminiscent of the two states of low brightness/hard spectrum and 
very high brightness/steep power law. If the QPOs are actually associated with an X-ray state it might not be a surprise not to see
all QPOs in one observation, like in the August 2004 X-ray flare, for example. 

Up to the middle of 2005 further QPOs from Sgr A$\sp *$ 
have been reported, both for the $\sim$2200 s (in the NIR) and the $\sim$1100 s range (X-rays). 
An overview has been  presented by Aschenbach (2006).
So far, there is no confirmation of the ultra-short periods around 220 s and below. This may be related to the limited 
time resolution of the NIR cameras. On the other hand    
the measuring of the 2200 s QPO is hampered by the fact that the flares are usually too short.

\subsection{Significance considerations}

Apart from the two X-ray flares with indications of QPO oscillations (Aschenbach et al. 2004) a third 
X-ray flare which occurred on August 31, 2004 shows a QPO signal (B\'elanger et al. 2005, B\'elanger et al.  
2006). Initially an average period of 21.4 minutes was quoted (Liu et al. 2006), which was later up-dated 
to 22.2 minutes after a Z$\sp 2$-periodogram analysis (B\'elanger et al. 2006).

There has been a lot of criticism and discussion about the significance of 
the X-ray QPOs, and I'm going to try to clarify the issue 
in the following by giving probability values. The most fundamental question is to what extent the observed 
power spectral density exceeds the level implied by photon counting noise. Photon counting noise is assumed to follow the 
Poisson distribution applied to an otherwise constant source. Although the statistics and related distributions are well 
known I simply use Monte Carlo simulations of noisy lightcurves and associated power density spectra 
for computing significances. I do not consider any other 
noise contribution like red noise because in case of Sgr A$\sp*$ we do not know the shape of the 
power density spectrum to simulate any realistic noise. It's a pure speculation to use power density 
spectra of AGN because we don't know to what part of the accretion disk they belong. A closer look to 
the power density spectra of microquasars with HFQPOs shows that they cannot be described by simple power 
laws.
For instance, the power density spectrum of GRS 1915+105 is curved, but can be approximated 
locally by a power law, the index of which is $\sim$-1 around the 168 Hz QPO, which corresponds to a fairly 
flat spectrum.

For reasons of comparison I re-analysed the flares of October 26, 2000 (Chandra) and  October 3, 2002 (XMM-Newton), the QPOs 
of which have been published (Aschenbach et al. 2004), and the August 31, 2004 flare in exactly the same way by binning the 
data by 120 s. The data of the August 31, 2004 flare have been simply copied from  Fig. 1 of B\'elanger et al. (2006). 
The data were decomposed in a Fourier series, the power density distribution was calculated and the maximum value of 
the power density was searched for. The periods 
of the Chandra flare and the early XMM-Newton flare, 
compared with the previously published results,  change slightly to  1151 s vs. 1117 s  and 1150 s vs. 1173 s.   
The corresponding power density can be reached by counting noise with a probability of 2.5$\times$10$\sp{-4}$ (Chandra) 
per trial  
and 7.1 $\times$10$\sp{-3}$ per trial (XMM-Newton), respectively. 
The equivalent analysis of the  August 31, 2004 flare reveals a period of 1320 s 
and a noise probability of 6.5 $\times$10$\sp{-6}$ per trial. The latter numbers agree with what B\'elanger et al. (2006)
have reported, i.e. 1330 s, 3.0 $\times$10$\sp{-6}$. It is absolutely essential to realize  that the power density 
can exceed the measured value at any of the involved Fourier frequencies, which means that the probability 
to find such a noise at just one frequency equals the probability per trial times the number of frequencies
(e.g. van der Klis 1989, Strohmayer 2001b) which is 
6.5 $\times$10$\sp{-6} \times 33 = 2.2\times $10$\sp{-4}$ for the August 31, 2004 flare, or a 3.7$~\sigma$ detection for a 
Gaussian probability  
distribution. So the probability that the alleged QPO power spectral density is due to white noise is a  factor of about 100 
higher than given by B\'elanger et al. (2006). 

The noise probability for the October 26, 2000 and  October 3, 2002 flare is larger and less significant per observation 
than for the August 31, 2004 flare (for the numbers see above).  
But the point made in the paper by Aschenbach et al. (2004) is that the enhanced power shows up at about the same 
frequency in these two observations. In numbers, with 71 frequencies (trials) covered, the probability 
to reach the observed powers in the two observations referred to 
 equals 71$\times$2.5$\times$10$\sp{-4} \times$ 7.1 $\times$10$\sp{-3}  = 1.3\times$10$\sp{-4}$, or a 3.8$~\sigma$ 
detection. So, the two quasi-periods of 1150 s and 1320 s appear to be of the same likelihood. 

Some more flares have recently been observed in the NIR and reported in the literature with periods of 743 s, 1043 s, 1217 s,  
1473 s and 2625 s (read off Figure 7 in Trippe et al. 2007) and 930$\pm$120 s (Meyer et al. 2006). 
Leaving aside the shortest and the longest period (743, 2625 s), 
which are close to some other frequencies reported before (Aschenbach et al. 2004), the  
remaining periods fall in a range between 930 s and 1320 s, the center of 
which is at 1162 s or 0.861 mHz with a width of $\pm$0.284 mHz FWHM. Numbers are obtained by averaging over the corresponding frequencies. 
Interpreting the relative bandwidth as coherence or quality factor, Q $\approx$ 3, which according to Remillard and McClintock (2005) 
would discriminate this QPO from 
a simple broad power peak. Furthermore, this value of Q is not inconsistent with what the microquasars HFQPOs  
show (Remillard \&\ McClintock 2006). 
For instance, the two highest frequency QPOs of GRS 1915+105 have Q$~\approx~$2--3 (c.f. Figure 11 
in Remillard \& McClintock 2006). 
The fact that we don't see the full frequency distribution may have to do with the limited statistics, 
such that the one or the other frequency dominates the band in a single observation. 
Furthermore, in a typical 10 ks observation the few hundred Hz QPO of a microquasar is sampled a few million times whereas 
the equivalent QPO of a supermassive black hole like Sgr A$\sp *$ is sampled just a few times. 
So the spread in frequency observed around $\sim$1100 s might be nothing else than the expression of the quality 
factor Q of a single QPO. 

\section{The Aschenbach effect or the 'humpy' frequency}

A couple of years ago I reported the discovery that very rapidly rotating black holes with spins of $\it a~>$~0.9953
show a  non-monotonic profile of the orbital velocity for Kepler motion (Aschenbach 2004).
This totally unexpected phenomenon shows up in the coordinate system of the 
{\it{Zero Angular Momentum Observer}} (ZAMO) also known as the {\it{Locally Non-Rotating Frame}} (LNRF). 
With decreasing radius the velocity rises initially, reaches a local peak, decreases with decreasing radius, runs through
a minimum and continues to rise again. Between the minimum and maximum the profile has a positive slope and it has the dimension
of a frequency, the value of which changes with the spin. I considered this frequency to be  a real physical frequency, 
which can be compared with the epicyclic
frequencies, for instance. The effect occurs only in a narrow region below a radius of 1.75 gravitational radii but above 
the innermost stable circular orbit. If this effect is actually the trigger mechanism to launch resonant oscillations 
the HFQPOs arise only in those black holes with a spin $\it a~>$~0.9953 suggesting a selection effect among black 
 holes as far as the occurrence of resonant HFQPOs is concerned.     
Z. Stuchl\'ik and his colleagues were so kind to call this effect the Aschenbach effect (Stuchl\'ik et al. 
2005). Because of the hump in the velocity profile they later called the associated frequency taken at the maximum of the profile slope the 
'humpy' frequency (Stuchl\'ik et al. 2006, Stuchl\'ik et al. 2007).

With the 'humpy' frequency a third frequency is provided to determine uniquely mass and spin of the black hole from 
one set of frequency measurements, involving the radial and vertical epicyclic frequencies. 
Originally calculated in the locally non-rotating frame (LNRF) using Boyer-Lindquist coordinates it was found 
that the 'humpy' frequency could match the radial epicyclic frequency at a radius, at which the frequencies of the vertical 
and radial epicyclic frequencies come in a 3:1 ratio. Applied to Sgr A$\sp *$ the mass and spin previously reported 
(Aschenbach 2004, Aschenbach 2006) could be confirmed, if the $\sim$700 s QPO is associated with the vertical epicyclic 
frequency and the $\sim$2100 s with the radial epicyclic frequency. It is unfortunate, that these two QPOs are still missing 
confirmation (but see Aschenbach 2006).
   
Recently Stuchl\'ik et al. (2006, 2007) have suggested that the 'humpy' frequency should be calculated in a coordinate independent 
form and for an observer at infinity. In fact, with the corresponding transformations the value of the 
'humpy' frequency changes but remains comparable in size with the epicyclic frequencies. The orbital radius at which 
the 'humpy' frequency occurs is in the vicinity of those radii at which the vertical and radial epicyclic frequencies 
come in ratios of 3:1, as before, or 4:1, where: "efficient triggering with frequencies of about the 'humpy' frequency 
could be expected" (Stuchl\'ik et al. 2007). 

They applied their model to the microquasar 
GRS 1915+105 and they can explain the four observed QPOs with frequencies of 41 Hz, 67 Hz, 113$\pm$5 Hz and 
166$\pm 5$ Hz. Note that both the 
first pair and the last pair of frequencies shows a ratio of 3:2. The 41 Hz oscillation is identified with the 
'humpy' frequency and the 67 Hz with the vertical epicyclic frequency. The next higher frequency at 113$\pm$5 Hz 
would correspond to the beat frequency, i.e. the sum of the 'humpy' and the radial epicyclic frequency, and the 
166$\pm 5$ Hz is the difference beat frequency, i.e. the difference between the vertical and radial epicyclic frequency. 
This model is the first model which explains consistently all four frequencies observed in GRS 1915+105. 

The 
model suggested by Aschenbach (2004), which was also based on the 'humpy' frequency but given in the LNRF and 
Boyer-Lindquist radial coordinate, explained just the two higher frequencies with a black hole mass 
of 18.1$\pm$0.4 solar masses, $\it a$~=~0.9962 and a resonance radius of 3.92 gravitational radii, 
whereas Stuchl\'ik et al. (2006) find 14.8 solar masses, $\it a$~=~0.9998 and a resonance 
radius of 1.29 gravitational radii. 
In general, the 'humpy' frequency model allows the determination of both the black hole mass and the spin 
from the measurement of a minimal of just three frequencies. It restricts the so-called high frequency 
resonant QPOs to extremely  fast rotating black holes and the oscillations are launched at the very inner edge of 
the otherwise stable parts of the accretion disk.      
It will be interesting to see how these results, in particular the spin, compare with those results determined 
by other methods, like X-ray continuum fitting, Fe-K line fitting and polarimetry. However, so far these latter 
methods agree on that the black holes 
are indeed Kerr black holes, but there is still a significant spread of $\it a$ among the individual models for the same object. 
At some time these models might be the ultimate test for the 'humpy' frequency model and vice versa.

\bigskip
\noindent
{\b DISCUSSION}
\bigskip

\noindent
{\b GENNADY BISNOVATYI-KOGAN:} Does the anomalous orbital velocity effect happen inside the ergosphere?
\bigskip

\noindent
{BERND ASCHENBACH:} Yes, it does, at least if the orbital motion is 
in the equatorial plane. The behaviour outside the equatorial plane has not been studied, 
yet, as far as I know.
\bigskip

\label{lastpage}
\end{document}